# Functionality and Protein-Water Interactions

J. C. Phillips

Dept. of Physics and Astronomy, Rutgers University, Piscataway, N. J., 08854-8019

## Abstract

The structures of proteins exhibit secondary elements composed of helices and loops. Comparison of several water-only hydrophobicity scales with the functionalities of two repeat proteins shows that these secondary elements possess water-induced medium-range order that is sometimes similar, but can also be complementary, to structural order. Study of these hitherto "phantom" order parameters promises far-reaching incremental improvements in the theory of protein dynamics. A by-product of the theory is an independent evaluation of the reliability of different hydrophobicity scales.

Proteins are polypeptide chains with side groups selected from a list of 20 amino acids (aa's). The sequences of these aa's are known for many proteins, while the corresponding structures have been determined for only a small fraction of these. So far the general problem of combining secondary structures to obtain the complete protein structure (protein folding) remains unsolved. Although much progress has been made, in recent years research has reached a level of exponentially diminishing returns, as one would expect from seeking a general solution to such a complex problem[1].

Instead of seeking a general solution to the problem of the inter-relations between sequence, structure and functionality, one can adopt a different and more selective approach, better suited to achieving steady (and even growing) incremental progress. The simplest protein structures are found in the repeat family, in which each protein consists of a series of repeated alpha helices connected by loops and short turns[2]. Within the family one can identify consensus side group sites, nearly always occupied by the same one or two aa's, even when there are variations in repeat lengths. Sets of these consensus sites are statistically powerful markers, and 10,000 to 20,000 repeats have been so identified[3]. Quantitative effects are multiplied in repeats, enabling the recognition of hidden mechanisms.

Protein structures are dominated by the effects of hydrophobic collapse, which optimizes aa packing for both stability and functionality. Hydrophilic side groups are concentrated on or near surfaces, while hydrophobic side groups cluster in the protein core. Hydrophobicity scales based on solvation transfer energies of each of the 20 aa's in water with the corresponding energies in an organic solvent scatter (20-30%), depending on the organic solvent used, and the functional hydrophobicity of a given aa is different in long helices from what it would be in isolation.

Pintar has used hydrophobic collapse itself to define multiple hydrophobicity scales $S$ based separately on buried depths of aa's at helical sites and all sites[4]. Two Pintar hydrophobicity scales are shown for the aa "alphabet" [H = Glycine = G, $CH_3$ = Alanine = A, $(CH_3)_2CH$ = Valine = V, etc] in Fig. 1. Zebende cleverly utilizied the concept of solvent accessible surface areas (SASA)[5]. By itself this is an old idea: one determines the SASA of each aa by Voronoi tessellation based on Van der Waals radii, and then fixes the surface area accessible to a 1.4 A spherical probe (water molecule). Now comes a highly original idea: the scale is based not on the average SASA seen by a given aa ($S(1)$), but rather on the way this SASA contracts with increasing N in a helical environment of 2N +1 aa's on a chain, $1 \le N \le 17$. (This corresponds to virtual folding of each helix centered on a given aa; systematic behavior associated with chain curvature emerges upon averaging overlapping SASA's over a large number of cases, 5526 ultra-high resolution helical fragments.) Each SASA contracts self-similarly, according to SASA(aa) = $S(0)(2N+1)^{-\gamma(aa)}$. One can use $-\gamma(aa)$ as an "ideal" hydrophobicity scale: note that it does not have dimensions of either energy or length, in fact, it is dimensionless (third line, Fig. 1).

Which of the three scales is best? An easy way to test the obtained values is to introduce helical and loop configuration coordinates for medium range (8-20 aa's) secondary structures. For each secondary unit in a given aa sequence one introduces first the average $\Psi(S) = <-\gamma(aa)>$. For repeats indexed by S one can define a second measure of hydrophobic stiffness or flexibility,

$$\Phi(S) = \Sigma[(\gamma_i(R(S)) - \gamma_i(R(S+1)))^2 + (\gamma_i(R(S)) - \gamma_i(R(S-1)))^2]/2M \qquad (1)$$

Here R denotes an amino acid and $\gamma_i(R)$ is its hydrophobicity. Consecutive repeats are aligned in the standard matrix tableau dictated by the consensus set; the sum is over matched helices of maximum length M, so that both $\Psi$ and $\Phi$ are normalized. Even in the context of adjacent (nearly parallel non-repeats, no consensus sites) $\alpha$ helices (such as H-bonded $\beta$ strands), $\Phi$ could be useful.

One can compare patterns of $\Psi$ and $\Phi$ for helices and loops with functionalities; in spite of their family structural similarities as spring-like coiled coils, the $\Psi$ and $\Phi$ patterns vary characteristically from one repeat protein to the next. We focus on two cases which provide powerful examples, one a simple case with important implications for oncogenic mutations (Fig. 2(a)), the other for measuring the accuracies of hydrophobicity scales (Fig. 2(b)). Like many other repeat proteins, both PR65/A (a scaffold 15-repeat, 90% helices, no loops, PR65/A (592 aa's)[6] and the 19-repeat importin $\beta$ (876 aa's)[7], have two helical arms (A and B)/repeat, forming "L cupped hand" or "chopstick" structures, suitable for differentiating "inside" from "outside", and for attaching to other proteins.

The overall structures of the ankyrin 12-repeat D34 (420 aa's)[8] and PR65/A are simpler than importin $\beta$, in the sense that their secondary $\alpha$ helices and loops are regularly repeated, while those of importin $\beta$ are somewhat irregular, but the consensus sites of importin $\beta$ resemble those of PR65/A. The $\Psi$(A,B) and $\Phi$(A,B) patterns of the helical arms of PR65/A, shown in Fig. 3, are strikingly different, and reflect different functionalities. Some of these differences are already obvious from the spatial structures[6-8], but many are phantoms that become obvious only when viewed in the "magic" light of medium range hydro(phobic/flexibl)ity $\Psi$ and $\Phi$ configuration coordinates.

Generally speaking, hydro(phil/phob)ic interactions with water soften/harden helices; the left (N) half of PR65/A (Fig. 3(a)) exhibits hard comb-like $\Psi$(B) arms and relatively soft $\Psi$(A) arms, with a hydro hinge (both arms below hydroneutral at 0.155) at the central repeat S = 8. The A arms are associated with the convex outer surface, while the B arms belong to the concave inner surface, which functions as the scaffolding support[6] for catalytic and regulatory domains. The regulatory subunit B56γ1 (itself a 16-repeat) attaches[9] to repeats 2-7. The catalytic subunit attaches to the B arms of repeats 11-15, which are associated with large oscillations in Ψ. Meanwhile the (A,B) hydrofragility Φ patterns (Fig. 3(b)) are quite different. Apart from softness near the N end, there is little structure between the A arms, but the B arms show distinct fragility peaks, at S = 13, 4, and 12 (in that order). One of the helical consensus sites in the B arm is 24, which is occupied by V in 11 out of the 15 repeats. This hydrophobic aa is missing from repeats 1 (N end), 4, 12 and 13, an essentially perfect correlation with the fragility peaks 13, 4, and 12. The marginal fragility of B arms of S = 8 and 9 correlates with the hydrohinge seen in Ψ in Fig. 3(a).

The 3 aa turns connecting the A and B arms form two distinct groups. Thus 10 of the turns are soft (strongly hydrophilic, $<\Psi> \sim 0.10$), while 5 are close to hydroneutral (($<\Psi> \sim 0.14$, repeats 3,4,8,10,14). The latter may promote stability of the marginally stable scaffolding structure. Four oncogenic site mutations associated with lung and colon tumors have been discussed in terms of stability of the dry structure[6], but here one can propose an alternative *in vitro* interpretation. Two of the mutations occur in helical sites, and these two (P(65)→S and L(101)→P) involve destabilizing decreases in hydrophobicity. The other two occur in the short turns, D(504)→G (repeat 13) and V(545)→A (repeat 14). The former stiffens the typically hydrophilic repeat 13 turn, while the latter softens the atypically nearly hydroneutral repeat 14 turn; both mutations regress Ψ for these turns towards the mean value (0.118) for all 15 turns. All of these changes favor more rapid production of the oncogenic protein, while disrupting its functionality. These correlations are invisible in the context of the spatial structure alone[6].

Transport of proteins from synthesis to functional sites is mediated by a multiplicity of repeat proteins, which typically load and unload their cargoes through large-scale conformational changes[10]. The binding function of 19-repeat importin β involves a 9 aa loop between the A and B arms of repeat 8. This loop is strongly acidic (Ψ = 0.093, deep in the hydrophilic aa tail, far below hydroneutrality at 0.155), but that is not all. While there is nothing special about Ψ(A,B) near repeat 8 (Fig. 4(a))), there is a striking dip/peak (hard/soft) bifurcation in hydrofragility $\Phi$(A,B) there (Fig. 4(b)). This is the largest A/B helical arm repeat asymmetry seen in our calculations, making it suitable for testing different hydrophobicity scales.

Fig. 5 focuses on $\Phi$(A,B) in the importin $\beta$ region centered on repeat 8. This figure shows subtle yet robust incremental trends. In (a) the two buried depth depth scales put the flexibility peak at repeat 7, instead of repeat 8, and the A arms $\Phi$ incorrectly cross the B arms for repeats 9 and 10. The mean depth does better than the helix on repeat 8 compared to repeat 7. Overall one can say that the (helix,mean) depth scales are (good,better). In (b) we see that the self-similar scale improves on the mean scale, with the peak at repeat 8 and the A arms below the B arms even for higher repeats (see also Fig. 3(b)). Thus the (mean,self-similar) scales are valued as (better,best).

In conclusion, these results show that hydrophobicity indeed has self-similar properties, not only on the scale of global domain networks[11], but also on the microscopic scale of individual aa's. By selectively studying incremental functional trends for (good,better,best) hydrophobicity scales, one can establish super-statistical valuations of phantom water-protein interactions that provide valuable insights into protein structural dynamics. (For example, it has been shown elsewhere that the "plastic" trends in m fragility slopes observed for the D34 repeats[8] in urea(wedge)-induced transitions are explained by decoupling of the A and B arms. This should be compared to the trends in related m fragility viscosity slopes at the glass transitions of simple hydrocarbon alcohols and saccharides[12].) These calculations were performed on a PC using only EXCEL spreadsheets and functions. History shows that from such humble beginnings sustained incremental progress can be expected. For example, pseudopotential theory began merely as a three-parameter empirical model for the energy bands of Si and Ge[13]. Because they are transparent, reliable, and transferable, pseudopotentials today dominate the citation indices of all combined physics disciplines[14].

# Figure Captions

Fig. 1.  Alphanumeric hydrophobicity tables.  Amino acids are coded by letters, and the three scales have been adjusted to span similar ranges with similar averages (hydroneutrality) near

0.155. The entries in order correspond to (1) helix depth[4], (2) mean depth[4] and (3) self-similar SASA[5]. Linear regression fits give standard deviations $\sigma_{12} = 0.012$ , $\sigma_{13} = 0.016$, and $\sigma_{23} = 0.018$, with most of the errors at large hydrophobicities. For ordinates in Figs. 3-5 tabulated numbers were multiplied by $10^3$.

Fig. 2 (a) Rod picture[7] of importin β (in color on line): the A (B) helical arms are in red (yellow). The cargo is IBB, and the acidic loop connecting 8(A,B) is in blue[7]. (b) Ribbon picture[6] of PR65/A.

Fig. 3. Ψ and Φ diagrams for the scaffold protein PR65/A (in color on line): A (B) helical arms are in blue diamonds (red squares), their average is in green triangles. Note that Φ(B) displays three peaks. The 2-7 peak binds a regulatory subunit, a catalytic subunit is bound to 11-15, and the central Φ peak is associated with the interface between these two subunits.

Fig. 4. Ψ and Φ diagrams for importin β (in color on line): A (B) helical arms are in blue diamonds (red squares).

Fig. 5. Comparisons of three scales near repeat 8 (in color on line). (a) Helix A, green triangles; helix B, violet crosses; mean A, blue diamonds, and mean B, red rectangles. (b) mean A, green triangles; mean B, violet crosses; self-similar A, blue diamonds, and self-similar B, red rectangles.

| A | 0.139 | C | 0.161 | D | 0.070 | E | 0.067 | F | 0.196 |
|---|-------|---|-------|---|-------|---|-------|---|-------|
|   | 0.156 |   | 0.179 |   | 0.077 |   | 0.068 |   | 0.219 |
|   | 0.157 |   | 0.246 |   | 0.087 |   | 0.094 |   | 0.218 |
| G | 0.142 | H | 0.110 | I | 0.235 | K | 0.058 | L | 0.209 |
|   | 0.102 |   | 0.122 |   | 0.246 |   | 0.064 |   | 0.217 |
|   | 0.156 |   | 0.152 |   | 0.222 |   | 0.069 |   | 0.197 |
| M | 0.197 | N | 0.081 | P | 0.097 | Q | 0.071 | R | 0.077 |
|   | 0.196 |   | 0.087 |   | 0.097 |   | 0.083 |   | 0.087 |
|   | 0.221 |   | 0.113 |   | 0.121 |   | 0.105 |   | 0.078 |
| S | 0.077 | T | 0.129 | V | 0.213 | W | 0.194 | Y | 0.157 |
|   | 0.100 |   | 0.117 |   | 0.225 |   | 0.197 |   | 0.169 |
|   | 0.100 |   | 0.135 |   | 0.238 |   | 0.174 |   | 0.222 |

Fig. 1.

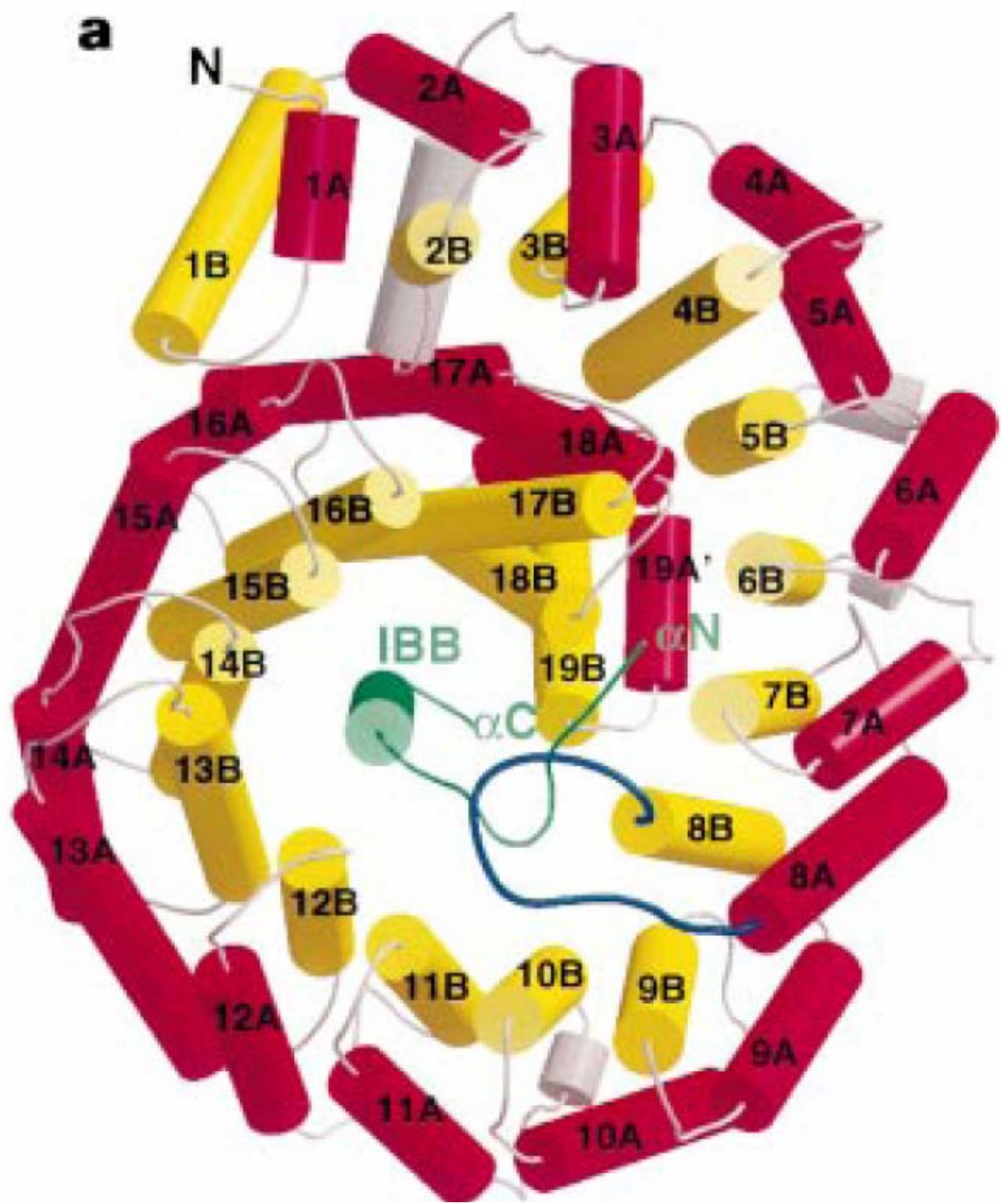

Fig. 2(a)

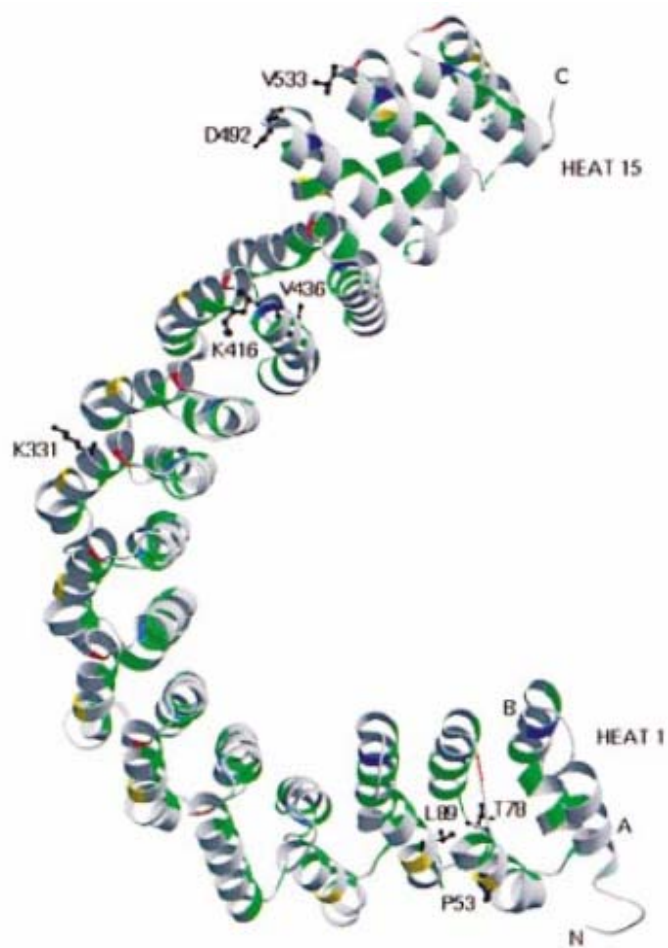

Fig. 2(b)

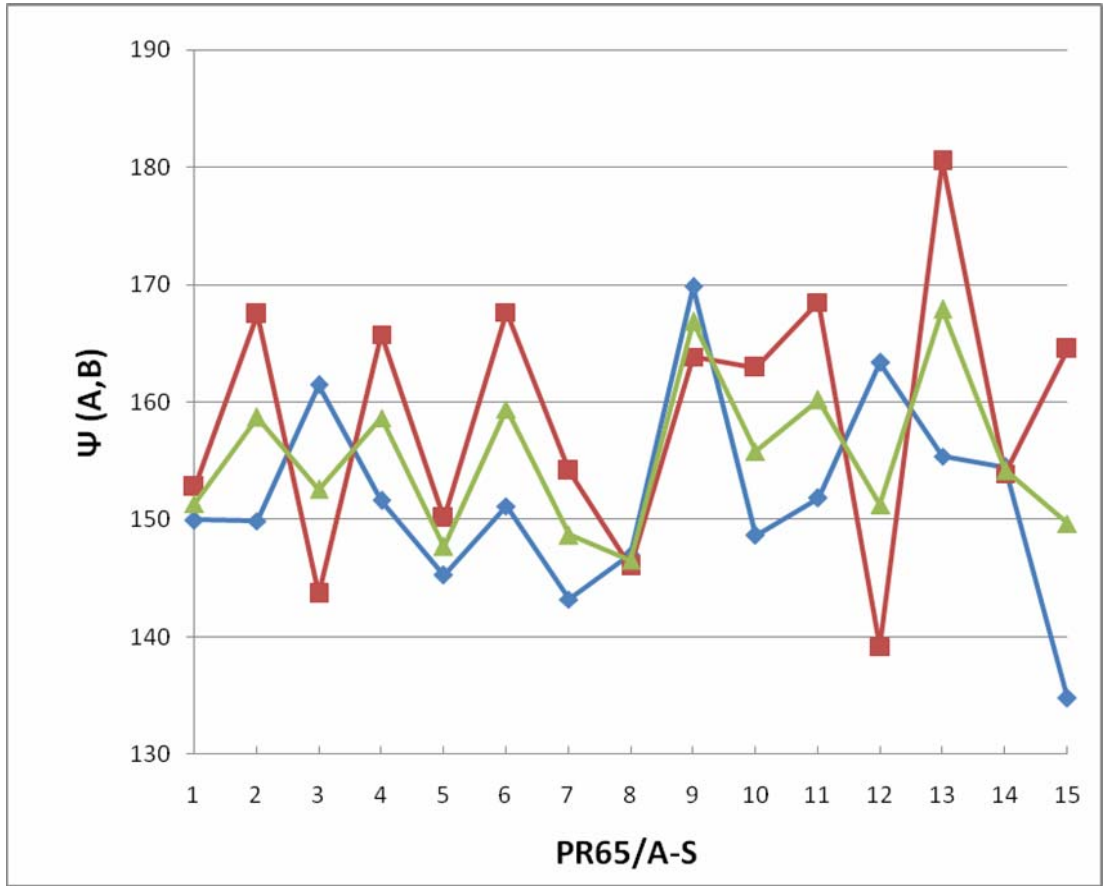

Fig. 3(a)

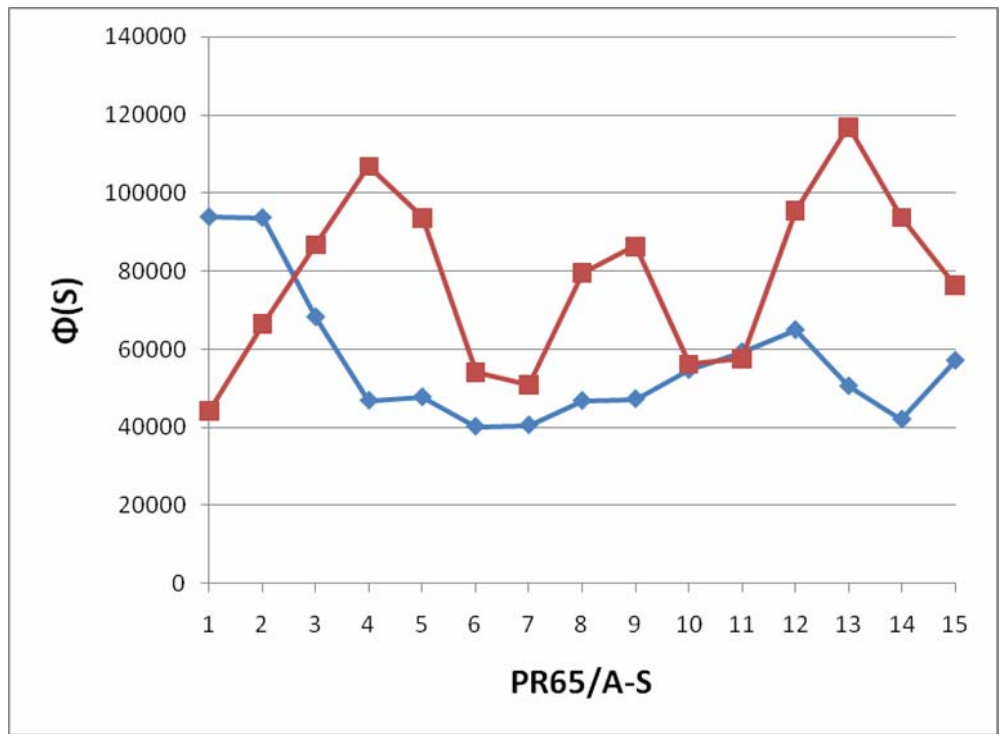

Fig. 3(b)

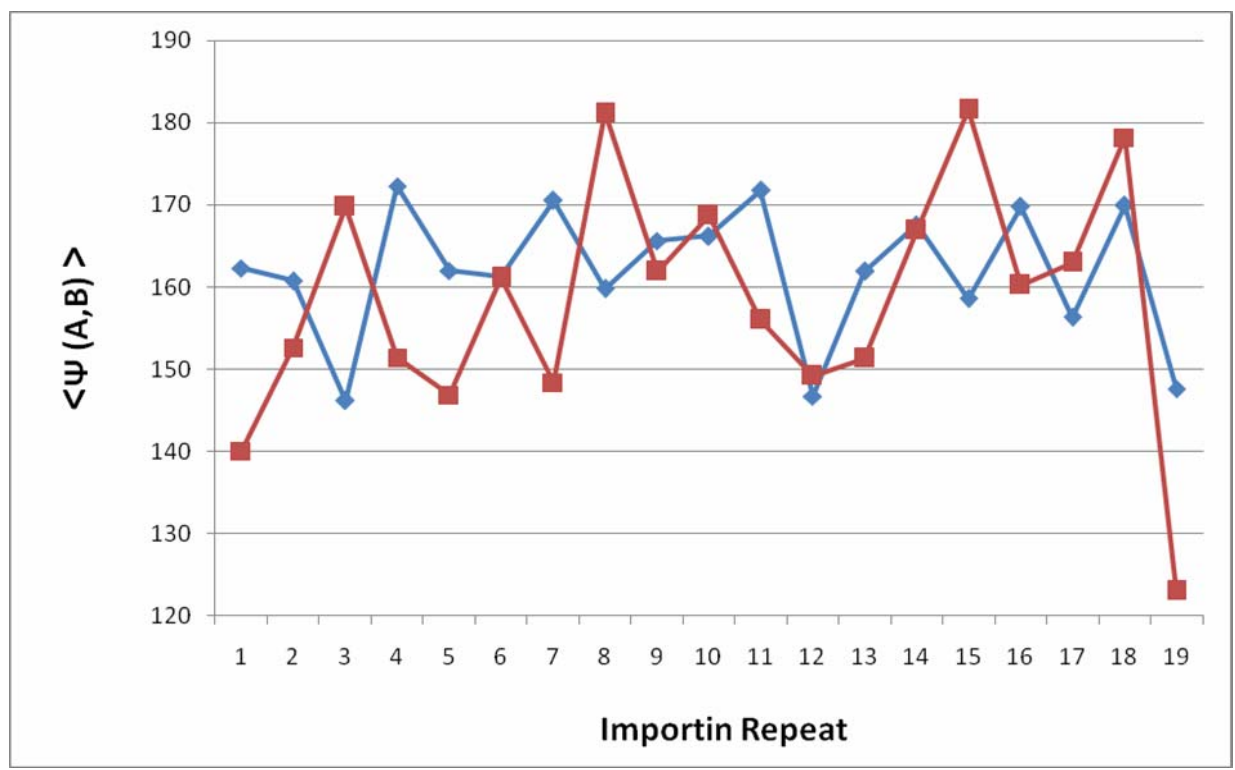

Fig. 4 (a)

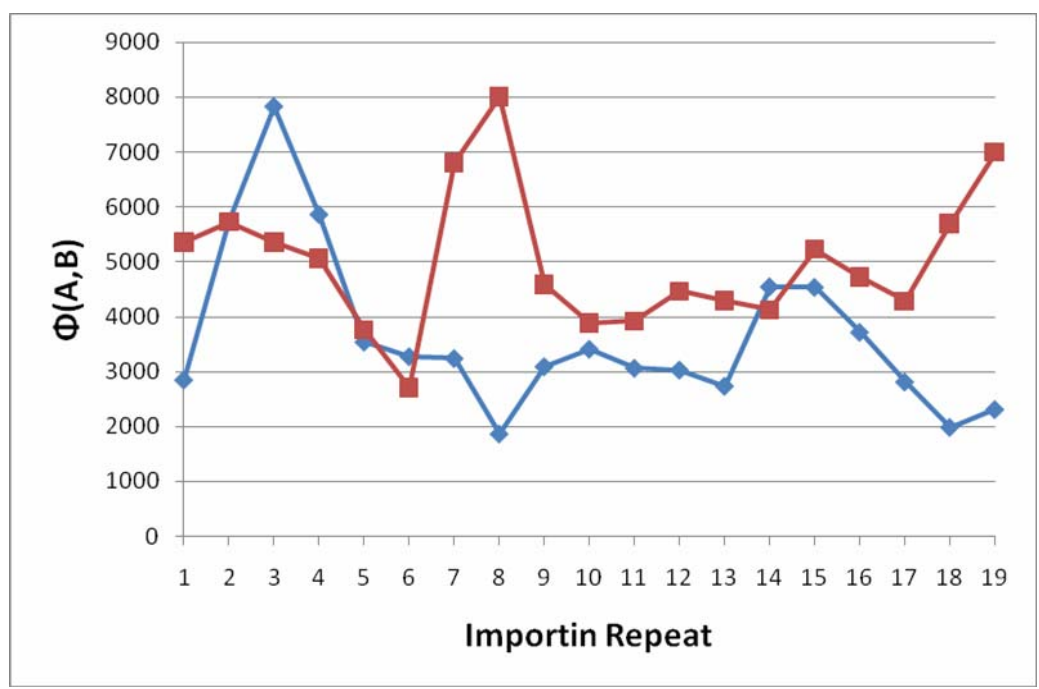

Fig. 4(b)

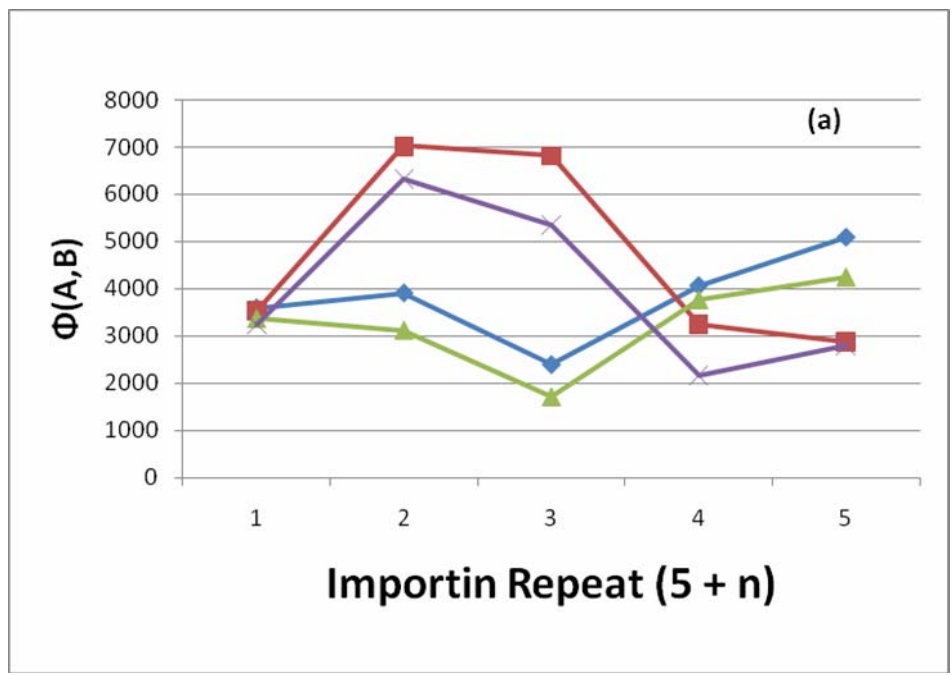

Fig. 5(a)

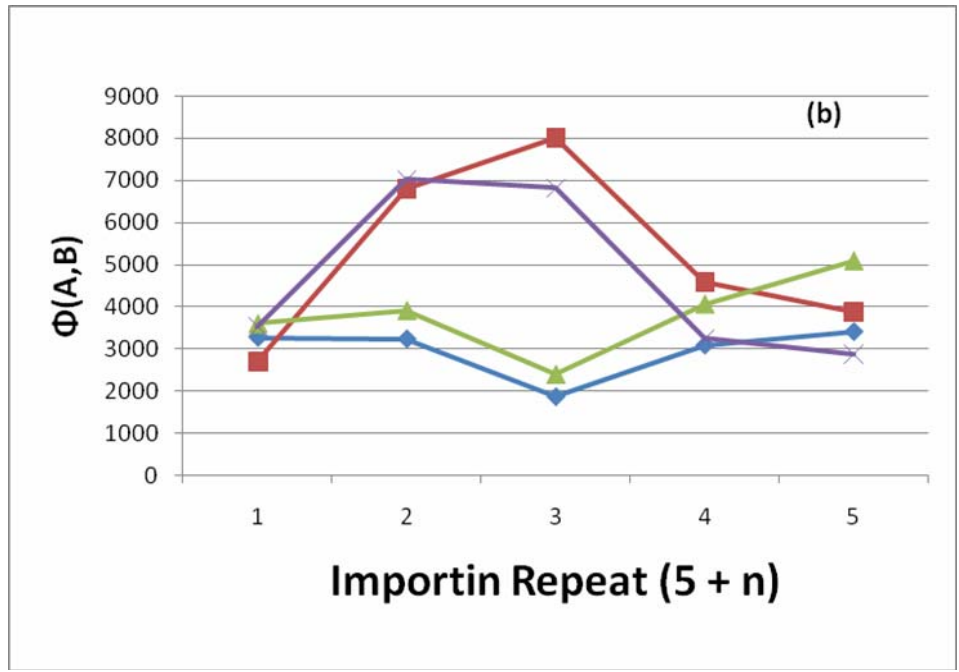

Fig. 5(b)

82.30.Rs 87.15.km 87.15.kr 87.15.H-